\documentclass[showpacs, twocolumn]{revtex4}
\usepackage{latexsym}
\usepackage{graphicx}
\usepackage{color}
\begin{document}
\title{Coexistence of anomalous field effect and mesoscopic conductance fluctuations in granular aluminium}
\author{J. Delahaye, T. Grenet and F. Gay}
%\and Second author\inst{2}% etc
%Institut Néel, CNRS-UJF, BP166, 38042 Grenoble, France
\address{Institut N\'eel, CNRS-UJF, BP 166, 38042 Grenoble, France}
\date{\today}
\begin{abstract}

We perform electrical field effect measurements at 4~K on insulating granular aluminium thin films. When the samples size is reduced below $\simeq 100\mu$m, reproducible and stable conductance fluctuations are seen as a function of the gate voltage. Our results suggest that these fluctuations reflect the incomplete self-averaging of largely distributed microscopic resistances. We also study the anomalous field effect (conductance dip) already known to exist in large samples and its slow conductance relaxation in the presence of the conductance fluctuations.
Within our measurements accuracy, the two phenomena appear to be independent of each other, like two additive
contributions to the conductance. We discuss the possible physical meaning of this independence and in particular
whether or not this observation is in favor of an electron glass interpretation of slow conductance anomaly
relaxations.

\end{abstract}

% end of PACS codes
%end of abstract
%
%\authorrunning{J. Delahaye et al}
%\titlerunning{Anomalous field effect and conductance fluctuations in granular aluminium}
\maketitle
\section{Introduction}
\label{intro}

An anomalous field effect in disordered insulators has been first observed in 1984 by Adkins et al. \cite{AdkinsJPC84}. In their experiment, an insulating discontinuous gold film was used as the conducting channel of a MOSFET structure. At low temperature, the conductance of the film was found to increase when the gate voltage was swept away from its cooling value, whatever the sweep direction. A few years later, Ovadyahu et al. observed a similar conductance dip in insulating indium-oxide (amorphous InO$_x$ and crystalline In$_2$O$_{3-x}$) thin films \cite{BenChorinPRB91}. A large number of experiments have been performed since then in the latter systems \cite{VakninPRB02}. The conductance was shown to decrease as a logarithm of time after a temperature quench or a sudden gate voltage change. The samples never reach a true equilibrium state even after one week of measurement. Qualitatively similar effects were reported in ultra thin films of Pb and Bi \cite{MartinezPRL97} and we published a thorough study of these phenomena in granular Al films \cite{GrenetEPJB03,GrenetEPJB07}. It is worth noting that doped semiconductors do not show a dip and that slow and glassy like electronic relaxations have been only rarely reported in there \cite{DonMonroePRL87,PopovicPRL06}.

A priori all these systems have different microstructures: the disorder is "granular" for discontinuous and granular metal films whereas it is "homogeneous" (oxygen vacancies and atomic position distribution) for indium-oxide films. Consequently, two different physical pictures have naturally emerged to explain the anomalous field effect and its long relaxation times.
\begin{itemize}
\item[-] In InO$_x$ films, it was shown \cite{VakninPRL98} that the typical relaxation time and the width of the conductance dip depend systematically on the carrier concentration $n$: the larger $n$, the longer the relaxation time and the wider the dip. Moreover, the relaxation time decreases rather sharply below $n\simeq 10^{20} cm^{-3}$. These results may explain why the anomalous field effect and its slow relaxation are observed in Anderson insulators with a high carrier concentration, like indium-oxide and granular Al, and not in doped semiconductors which usually have much smaller $n$ \cite{VakninPRL98}. They also suggest that such phenomena reflect the existence of an electron glass at low temperature (we call this hypothesis the intrinsic one). The electron glass existence was theoretically predicted for disordered insulators in 1982 by different authors \cite{DaviesPRL82,GrunewaldJPC82,PollakSEM82}. The combined effects of disorder and unscreened interactions between localized electrons should give rise to a glassy dynamics and correlated motion of electrons at low temperature. The experimental findings of slow conductance relaxations have motivated several theoretical works on the electron glass problem \cite{YuPRL99,GarridoPRB99,TsigankovPRB03,GrempelEL04,MullerPRL04,LebanonPRB05}.
\item[-] In granular systems, a different explanation was proposed. It was shown that the conductance dip and its slow relaxation could be qualitatively accounted for by a slow polarization of the dielectric material around the metallic grains \cite{GrenetEPJB07}. This idea was first introduced by Adkins et al. \cite{AdkinsJPC84} and developed later by Cavicchi and Silsbee \cite{CavicchiPRB88} for the interpretation of capacitive measurements on granular films. We call this hypothesis the extrinsic one since according to it, the non equilibrium effects don't come from the electrons of the metallic grains themselves. Other observations may be in favor of an extrinsic interpretation. Features qualitatively similar to field-gated conductance measurements were found at very low temperature in the dielectric response of amorphous materials to large DC electric fields \cite{SalvinoPRL94}. It was also shown that bias voltage changes can induce at room temperature non exponential relaxations and memory effects in the electronic properties of Al-AlO$_{x}$-Al planar tunnel junctions, the effects being attributed to the metastability of interface states \cite{NesbittPRB07}. Interestingly enough, such tunnel junctions are present in our granular Al thin films \cite{GrenetEPJB07}. Last, we would like to mention that an extrinsic type of scenario has been recently suggested for indium-oxide films \cite{BurinCondmat07}.
\end{itemize}
The striking similarity between the different experimental results strongly suggests that the same physical process is involved for all the systems. We have already discussed in Ref.~\cite{GrenetEPJB07} why we believe that the two hypotheses (extrinsic versus intrinsic) could be applied to both granular and homogeneous systems. Up to now and to our point of view, none of the hypothesis has been undoubtedly settled by the experiments. The logarithmic time dependence of the conductance and the simple aging relaxation laws observed after gate voltage changes can be interpreted as the response of independent degrees of freedom with relaxation times $\tau_i$ such as $\ln\tau_i$ has a flat distribution over the time scale of the experiments \cite{VakninPRB00,GrenetEPJB07}. Such an interpretation is obviously independent of the nature of these degrees of freedom (correlated electrons hops, two level systems charge configurations, etc.~\cite{KozubCondmat08}). One important difficulty in order to rule out or to establish one hypothesis comes from the absence of quantitative and indisputable predictions directly comparable to experimental data. Whether the intrinsic (electron glass) hypothesis is the correct one or not is of prime interest since the long response times measured (more than hours and days) may constitute a direct experimental evidence of its existence.

To shed light on the actual origin of these slow relaxation phenomena, we choose to perform field effects measurements on small size granular Al thin films. By small size, we mean small enough to observe reproducible gate voltage induced conductance fluctuations and to study if the conductance dip is modified by these fluctuations. In diffusive metals of mesoscopic size, the conductance was shown to be sensitive to details of disorder and even to the motion of single impurities (see Ref.~\cite{WashburnRPP92} for a review). Changing the gate voltage or the magnetic field modifies the quantum interference effects between elastic diffusion centers responsible for weak localization corrections. At low temperature, this gives rise to the well known Universal Conductance Fluctuations, which are of the order of the quantum of conductance $e^2/h$ for samples having sizes equal or smaller than the coherence length. Gate voltage or magnetic field conductance fluctuations pattern could therefore be considered as a finger-print of the sample specific realization of disorder.

A similar finger-print exists also at low temperature for hopping disordered systems. Due to the large (exponential) distribution of electron hops probabilities between pairs of localized states, the conductance of small size samples is dominated by a few critical hops or microscopic resistances \cite{ShklovskiiEfros84}. The gate voltage and the magnetic field could change the values and the positions of the dominant resistances, giving rise to conductance fluctuations of quantum \cite{ShklovskiiSpivak91,FengPRL91} and/or geometrical origin \cite{LeePRL84}. The typical distance between the dominant resistances defines the length scale above which the conductance starts to self-average \cite{ShklovskiiEfros84,FengPRL91}. This length scale could be orders of magnitude larger than the phase coherence length (the hopping length in the variable range hopping regime) when the spread of hops resistances is very large. Indeed, conductance fluctuations have been observed in samples of sub-micrometer \cite{FowlerHT91,LadieuJPI93,HofheinzEPJB06} micrometer \cite{AronzonPA97,SavchenkoSM91} and millimeter sizes \cite{MillikenPRL90,PopovicPRB90}. Telegraphic noise of individual fluctuators was also identified \cite{SavchenkoSM91,CobdenPRL92,HofheinzEPJB06}. Depending on the system parameters and the temperature, the amplitude of the fluctuations can be as large as the conductance itself. As we will discuss in more details in Sect.~\ref{Interpretation}, similar gate voltage conductance fluctuations are expected in our insulating granular Al films. Since the fluctuations pattern obtained by scanning the gate voltage is believed to reflect the spatial and energetic distribution of microscopic resistances, it is especially interesting to see how it is influenced by the field effect anomaly and its slow relaxation. We published preliminary results in Ref.~\cite{GrenetTrieste03,GrenetEPJB07} and similar investigations were reported very recently on In$_2$O$_{3-x}$ films \cite{OrylanchikPRB07}.

In Sect.~\ref{Experiment} we will present samples elaboration and measurement techniques. Sect.~\ref{Fluctuations} is devoted to the study of reproducible conductance fluctuations in samples of micrometer size. The data are shown to be in agreement with a percolation model applied to strongly inhomogeneous media.  In Sect.~\ref{CondMin}, the conductance fluctuations are studied in relation to the anomalous field effect (the conductance dip) and slow relaxation phenomena. The conductance dip formation and its slow response to gate voltage changes are shown to have no influence on the fluctuations pattern. Conversely, the existence of conductance fluctuations does not modify the conductance dip parameters.

\section{Samples and experimentals}
\label{Experiment}

\subsection{Samples elaboration}
\label{Samples}

\begin{figure}[h]
\begin{center}\includegraphics[width=7cm]{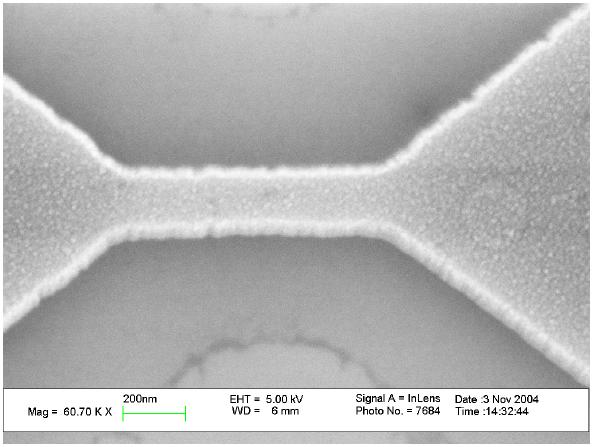}\end{center}
\caption{Scanning electron microscope picture of a granular Al channel made by electron lithography. The line is 180~nm wide and 820~nm long.}
\label{Picture}
\end{figure}

The samples used in this experiment are MOSFET devices in which the conducting channels are made of insulating granular Al thin films. To obtain them, granular Al is evaporated on top of a heavily doped Si wafer (the "gate") covered with a 100~nm thick SiO$_2$ insulating layer (the "gate insulator"). The evaporation was done either through a hand made mask (HM samples) or through a resist mask patterned by classical electronic lithography techniques (EL samples, see Fig.~\ref{Picture}). In this last case, 20~nm thick Au contact pads are evaporated prior to resist mask fabrication, whereas for hand made masks, Al contact pads could be evaporated on top of granular Al films without opening the vacuum chamber of the evaporator.
%This last point appears to be crucial in order to get linear I-V behavior at small voltages.

Granular Al films are deposited by e-beam evaporation of Al under a controlled pressure of oxygen. We focus here on samples lying on the insulating side but still close to the metal-insulator transition. The electrical resistance per square $R_{\Box}$ at 4~K is tuned between a few M$\Omega$  to a few G$\Omega$  by slightly increasing the oxygen pressure. Typical parameters are an Al evaporation rate of 2\AA/s and an oxygen pressure around $2.10^{-5}$~mbar. The film thickness is 20~nm for hand made samples and 40~nm for samples made by electronic lithography. The microstructure of the films was already discussed in Ref~\cite{GrenetEPJB07}. They are believed to consist of Al grains with typical diameters of 5~nm separated by thin insulating AlO$_x$ barriers.

\subsection{Electrical measurements}
\label{Electrical}

The electrical resistance of the granular Al channels was measured in a two contacts configuration. The sample was DC or AC voltage biased and the resulting current was measured through a home made or Femto DLPCA-200 current amplifier. A DC voltage was applied to the gate and could be swept between -30~V and 30~V. No current leak through the gate insulator was detectable in this voltage range. DC voltage sources for the gate and the bias were Yokogawa 7651 or AOIP SN830. AC measurements (frequency between 10~Hz to 200~Hz, depending on the sample resistance) were done with Signal Recovery 7265 or Stanford SR810 Lock In amplifiers.

All the measurements presented below have been done at low bias voltages, in the linear regime of the I-V curves: typical bias voltage is 50~mV for a 50~$\mu$m long channel.

\begin{figure}[h]
\begin{center}\includegraphics[width=8cm]{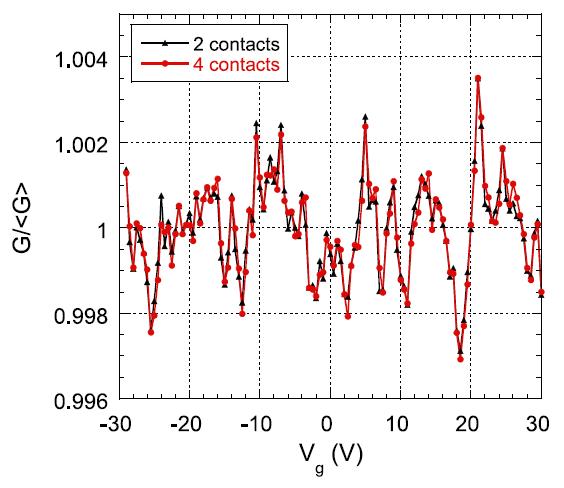}\end{center}
\caption{Normalized conductance measured as a function of the gate voltage V$_g$ for a channel $20\mu m$ long and $30\mu m$ wide ($20\mu m\times30\mu m$) in a 2-contacts (triangles, average of 20 V$_g$ sweeps) and 4-contacts (circles, average of 5 V$_g$ sweeps) configurations. HM sample, $R_{\Box}=2$~M$\Omega$ , $T=4.2$~K.}
\label{TestLeads}
\end{figure}

Some samples made by electronic lithography displayed pronounced non linear I-V effects for unexpectedly low voltages compared with hand made samples of similar sizes. They were not used in this study. For hand made samples, the absence of contacts effects in the measured conductance was carefully checked. First, I-V curves were measured for samples of different lengths and it was checked that they depend only on $R_{\Box}$ and the electric field $E$ across the granular Al channels. Second, for one not too resistive sample ($R_{\Box}=2$M$\Omega$) $20\mu m$ long and $30\mu m$ wide, we compared 2 and 4 contacts measurements. The measured conductances were very close (the $\simeq1\%$ difference is of the order of the instruments precision) and the gate voltage fluctuations patterns to be discussed in this paper were practically indistinguishable (see Fig.~\ref{TestLeads}).

\section{Conductance fluctuations in small samples}
\label{Fluctuations}

\subsection{A reproducible conductance fluctuations pattern}
\label{Reproducible}

When the samples size falls below $\approx 100\mu m$, clear and well resolved fluctuations of the electrical conductance are visible as a function of the gate voltage at 4~K. The typical relative amplitude is 0.1\% for a $50\mu m\times 50\mu m$ sample.

The fluctuations pattern does not depend on the measurement conditions. It is independent of the gate voltage sweep parameters (sweep rate between $8~s/V$ to $160~s/V$, sweep direction, etc.), of the instruments used (gate and bias voltage sources, current amplifiers) and of the working frequency (DC or AC measurements). Indeed, in addition to the test of Fig~\ref{TestLeads}, the dependence of the fluctuations amplitude with the square root of the granular Al channel area (see Fig.~\ref{OscilTaille}) makes up an indirect confirmation that the pattern actually comes from the channel itself and not from the contacts.

\begin{figure}[h]
\begin{center}\includegraphics[width=8cm]{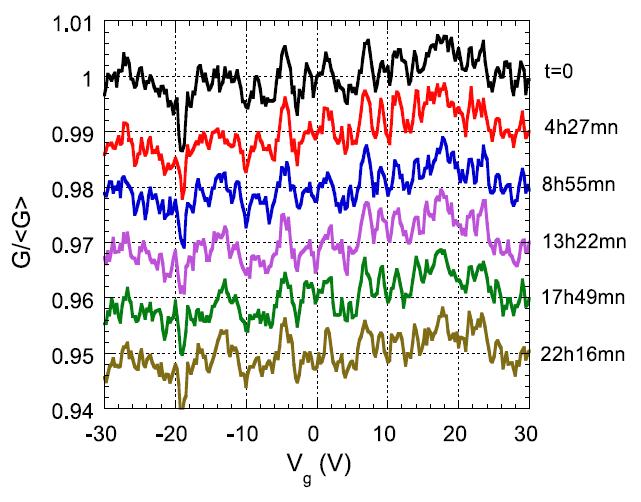}\end{center}
\caption{Normalized conductance measured over one day for a channel with a constriction of $340nm\times 230nm$ (EL sample, $R_{\Box}=3$~M$\Omega$ , $T=4.2$~K). The first gate voltage sweep (curve labeled $t=0$) was started less than one minute after a quench from 10~K down to 4.2~K. The other curves have been shifted for clarity.}
\label{J2Trempe}
\end{figure}

\begin{figure}[h]
\begin{center}\includegraphics[width=8cm]{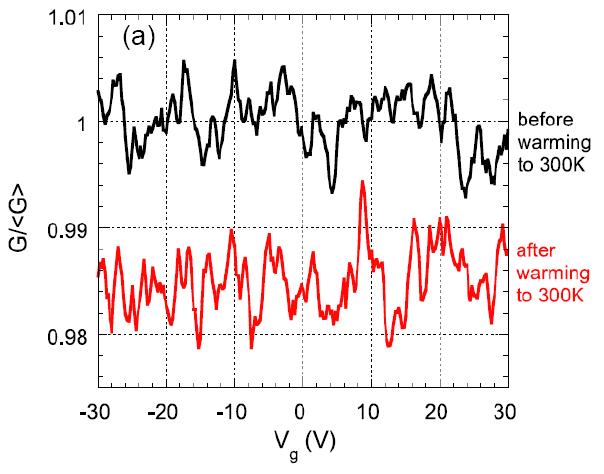}\end{center}
\begin{center}\includegraphics[width=8cm]{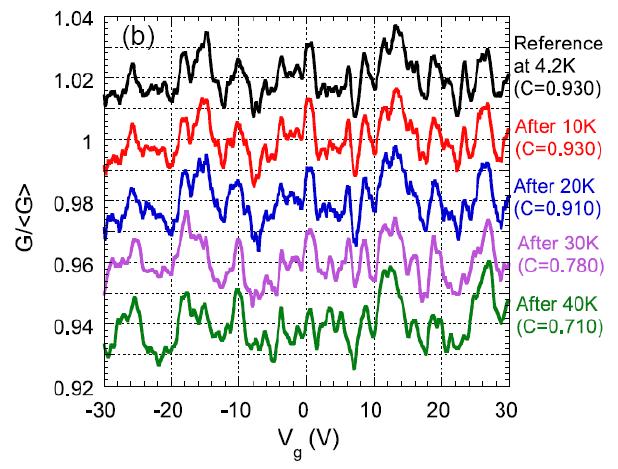}\end{center}
\caption{(a) Normalized conductance for a channel with a constriction of $340nm\times 230nm$ (EL sample, $R_{\Box}=3$~M$\Omega$). The upper curve corresponds to the fluctuations pattern at 4.2~K after a first cool down. The lower curve (shifted for clarity) corresponds to the new pattern obtained after the sample was warmed up to room temperature and cooled again at 4.2~K. The correlation coefficient C between the two curves is 0.19 while it is around 0.92 between successive sweeps at 4.2~K.
(b)~Normalized conductance for a channel with a constriction of $1.8\mu m\times 130nm$ (EL sample, $R_{\Box}=3$~M$\Omega$). From top to bottom: the initial fluctuations pattern at 4.2~K; the 4.2~K fluctuations patterns after  10~mn excursions at 10~K, 20~K, 30~K and 40~K. The correlation coefficients C indicated on the figure have been calculated between 4.2~K sweeps taken before and after the temperature excursions. The curves have been shifted for clarity.}
\label{J2CorT}

\end{figure}

As we can see in Fig.~\ref{J2Trempe}, the fluctuations pattern remains stable for one day after a quench at 4.2~K. Some noise is always superimposed on the fluctuations pattern that can be reduced by averaging successive sweeps. A careful check of the long-term stability often reveals a small decrease of the correlation coefficient C between successively measured patterns \cite{CorCoeff}, reflecting minor changes in the details of the fluctuations. If the sample is warmed up to room temperature and quenched again, the pattern is completely different (see Fig.~\ref{J2CorT}a). A more detailed study indicates that the temperature has to be raised above $\simeq30$~K in order to change significantly the fluctuations (see Fig.~\ref{J2CorT}b).

The distribution of the conductance values $G$ is roughly described by a Gaussian function. In disordered insulators and when the conductance fluctuations are large, asymmetries in the distribution of $\log G$ could give important information about the geometry of the critical resistance network \cite{OrlovSSC89a,HughesPRB96}. For example, a 1D chain will result in a tail towards low conductance values. But even in the sample with a narrow line 100~nm wide, we couldn't find any systematic asymmetry in the conductance distribution. This is probably due to the fact that our fluctuations are small (1\% or less of the conductance value).

\begin{figure}[h]
\begin{center}\includegraphics[width=8cm]{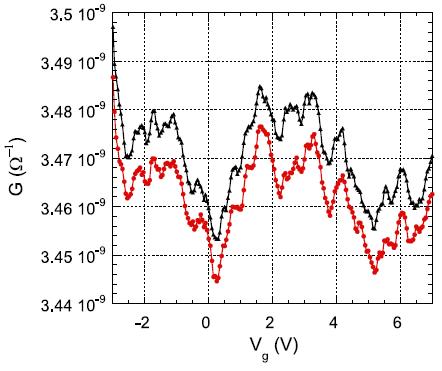}\end{center}
\caption{Conductance for a channel of $20\mu m\times 50\mu m$ (HM sample, $R_{\Box}=610$~M$\Omega$, $T=4.2$~K). The upper curve (triangles) is the average of 12 successive sweeps, and the bottom one (circles), shifted for clarity, the average of 9 sweeps. Small but well-reproducible structures are visible on a gate voltage range of  0.5~V. }
\label{QQNbrePoints}
\end{figure}

We characterized the fluctuations pattern by its root mean square (rms) amplitude $\sigma$ (standard deviation of $\{G(V_g)_i\}$ data). The noise contribution could be subtracted either by working out the average of many scans or by estimating its rms contribution (the variance of the difference between two scans gives an estimate of twice the variance of the noise). In order for $\sigma$ to be a well defined number, we used V$_g$ scans with at least one point every 0.5~V and over a range of $40-60$~V. For a given channel, $\sigma$ is found to vary by about 10\% between different coolings. Fourier Transforms of the fluctuations pattern does not reveal any V$_g$ periodicity and the autocorrelation function falls down to half its value for  $V_g=2-5$~V. A zoom of the fluctuations on a reduced V$_g$ scale is shown in Fig.~\ref{QQNbrePoints}. We can see that well-reproducible structures of only 0.5~V wide are also present.

\begin{figure}[h]
\begin{center}\includegraphics[width=8cm]{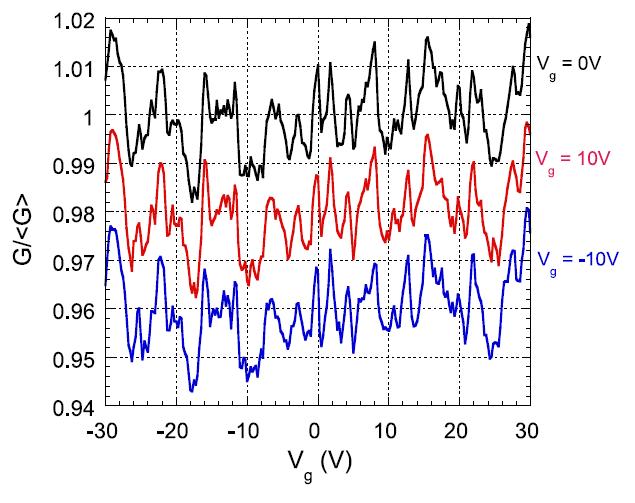}\end{center}
\caption{Normalized conductance for a channel with a constriction of $1.8\mu m\times 130nm$ (EL sample, $R_{\Box}=3$~M$\Omega$, $T=4.2$~K). The upper curve is the average of 9 sweeps taken after a long-time equilibrium under $V_g=0$~V. The middle curve is the average of 13 sweeps measured after the equilibrium gate voltage was changed to 10~V. And the bottom curve is the average of 7 sweeps taken after the equilibrium gate voltage was changed to -10~V. The duration of each sweep was 240~s, with 1 hour between two successive sweeps. Middle and bottom curves have been shifted for clarity.}
\label{J2ChgtVg}
\end{figure}

The pattern is also independent of the equilibrium gate voltage which is maintained between the sweeps (see Fig.~\ref{J2ChgtVg}). We will come back in more details to this observation in Sect.~\ref{CondMin}.

\subsection{General picture of the electrical conduction in our granular Al films}
\label{Interpretation}

We now have to wonder how to understand these conductance fluctuations and the features mentioned above. In particular, how could they be compared to the conductance fluctuations observed in "homogeneous" disordered insulators, like indium oxide?

As it was mentioned in Sec.~\ref{Experiment}, our insulating granular Al thin films consist of Al grains about 5nm in diameter separated by AlO$_x$ tunneling barriers. One important energy scale is the charging energy of an individual grain. Assuming a tunneling barrier thickness of 1~nm and a dielectric constant close to that of Al$_2$O$_3$, we get a charging energy of about 150~K. It is much larger than the thermal energy at 4~K and Coulomb blockade effects are thus important at low temperature in our system. The room temperature R$_{\Box}$ of the films spans from 30 to 80~k$\Omega$ (HM samples, 20nm thick). It should give a lower bound value for the dominant microscopic tunnel resistances \cite{note2}. Even if such values are not much larger than the quantum resistance $h/e^2\simeq 26$~k$\Omega$, we will neglect quantum fluctuations of charge on the grains in the following discussion.

In a perfectly regular 2D-array of metallic grains (i.e. an array with unique values of inter-grains capacitances C, gate-to-grains capacitances C$_g$ and tunnel resistances between adjacent grains), a global gate voltage V$_g$ induces the same polarization charge in each grain $Q_g=C_gV_g$. A naive picture suggests that at low temperature, such a V$_g$ change gives rise to a large and periodic modulation of the array conductance with a period of $e/C_g$, similar to what is observed in Single Electron Transistors or SETs. This is clearly not what we are observing here: relative conductance fluctuations are of only few percent or smaller. But as we recall now, this naive picture neglects important sources of potential disorder in the system \cite{AdkinsJPC84,AdkinsMIT95}.

The first source of disorder is the inevitable presence of grains size and tunnel barriers thickness distributions, which unfortunately we cannot measure easily. They would result in distributions of charging energies, gate-to-grains capacitances C$_g$ and tunnel resistances. But even if a wide distribution of C$_g$ will smooth the Coulomb oscillations for gate voltage larger than $\approx~e/C_g$, a large gate voltage conductance modulation should still be present at $V_g=0$ (grains in phase) which is again not observed in our films.

Another important source of random potential comes from the presence of charged impurities or defects trapped in the dielectric environment of the grains (tunneling barrier, substrate and natural oxide layer covering the film). Such trapped charges polarize the metallic grains and induce so called offset charges. The offset charges Q$_0$ are not quantized and could be larger than e. They can be static or dynamic. When a metallic island is connected to bulk electrodes with two tunnel junctions (SET), the current through the device at low bias (in the Coulomb blockade regime) was shown to be very sensitive to tiny changes in the island potential corresponding to offset charges of 10$^{-3}e$ or below \cite{BouchiatThesis97,LikharevNMT03}. The existence and the dynamics of offset charges were clearly identified in metal-based transistors made by electronic lithography techniques (typical island size is 1~$\mu$m) \cite{BouchiatThesis97,ZimmerliAPL92,ZorinPRB96,ZimmermanPRB97,KrupeninJLTP00,HuberReview01}, in scanning tunneling spectroscopy measurements on individual and oxidized metallic grains of nanometer size \cite{BentumPRL88,WilkinsPRL89} and in planar metal-based tunnel junctions \cite{RogersPRL84,McCarthyAPL99}. In our experiments, the granular Al films are deposited on a thermally grown SiO$_2$ layer 100~nm thick. We think that, in agreement with results on Al oxidation \cite{TanPRB05}, our AlO$_x$ dielectric layer around the metallic grains is not stoichiometric with Al$_2$O$_3$ and contains a large concentration of oxygen vacancies. This layer certainly constitutes the main source of potential disorder in our system. The role of interface states in the electronic properties of Al-AlO$_x$Al tunnel junctions was also emphasized in Ref. \cite{NesbittPRB07}. Capacitive studies on standard Si-SiO$_2$ wafers covered with an electron-gun evaporated Al layer, indicate typical charge densities about $10^{12}$e.cm$^{-2}$, i.e. about one charge every 10~nm (mainly interface states) \cite{SerretThesis02}.

One last source of disorder for the potential of the grains was considered in Ref.~\cite{CuevasPRL93,VergesPRB97}. When the grains are very small, surface effects (irregular shapes and sizes) contribute to random changes in the grains potential that are much larger than the mean energy level spacing $\Delta$, and that could be even larger than the charging energy $E_C$. For metallic grains of 5~nm in diameter, Cuevas et al. \cite{CuevasPRL93} estimate that this effect could by itself ionize more than half of the grains. In our case, by taking the approximate formula $\Delta=1/n(E_F)d^3$ (d is the diameter of the grains) and the density of states of bulk Al, we get that $\Delta$ is about 4~K.

All these effects together result in a large distribution of offset charges that could exceed many times the electron charge $e$. Such large offset charges distribution is usually unstable with respect to single electron tunneling between the grains. They are (partially) compensated by quantized electrons hops from other grains and/or from the electrodes. Theoretical studies on regular arrays of tunnel junctions have shown that the offset charges configuration that minimizes the electrostatic energy depends on the ratio $C/C_g$ \cite{JohanssonPRB00,KaplanPRB03}. In 1D and when $C_g\ll C$, a charge placed on one island is screened on a distance $(C/C_g)^{1/2}$ called the soliton length (in units of the array step-size) \cite{AverinLikharev91}. When $C\ll C_g$ (short screening length limit), it is usually assumed that the system ends with a random and uniform distribution of offset charges between $-e/2$ and $e/2$ \cite{CavicchiPRB88,MiddletonPRL93}. But in the opposite limit ($C\gg C_g$), numerical simulations on 1D arrays have found that starting from a random offset charge distribution, the interactions between electrons on distant grains induce energy favorable electron tunneling between the grains that smooth the potential \cite{JohanssonPRB00}. Similar simulations were also done on 2D-arrays smaller than the soliton length \cite{KaplanPRB03}. After electrostatic energy minimization, the density of states of single electron addition energies has a Coulomb gap equivalent to what is observed in homogeneous disordered systems (but with a different energy dependence). Moreover, the linear conductance at very low bias voltages ($eV\ll k_BT$) is thermally activated. The activation energy is about $0.1e^2/C$ and changes in the offset charges configuration induce activation energy fluctuations about 1/3rd of this value (these quantities are almost independent of the size of the system). In our granular Al thin films, a rough estimate gives $C=2.10^{-18}$~F and $C_g=7.10^{-21}$~F (soliton length $(C/C_g)^{1/2} = 20$).  Thus, electronic correlations between the grains may play an important role, even if the soliton length (about 100~nm) is still smaller than the system size. It is worth noting that the understanding of the electronic properties of granular systems is still an active theoretical field \cite{BeloborodovRMP07} in spite of decades of research.

For simplicity, we will consider now that we are let with a random potential disorder with a homogeneous distribution of offset charges on the different grains (we neglect electronic correlations). Such a distribution naturally explains why large Coulomb oscillations are not observed as a function of V$_g$ in macroscopic samples \cite{AdkinsMIT95}. It also implies (in the regime $k_BT\ll E_C$) an exponentially large distribution of hopping probabilities between the grains (a distribution in the barrier thicknesses will have the same effect)\cite{MullerPRB02}. Therefore, percolation theory results obtained for strongly inhomogeneous media \cite{ShklovskiiEfros84} should also describe the conductance of our granular Al films. According to this theory, our V$_g$ fluctuations come from an incomplete self-averaging of the conductance in small size samples ("mesoscopic" fluctuations). Like in homogeneous hopping systems, the conductance is dominated by a small number of critical resistances (which are here single electron transistors, possibly asymmetric) with a typical distance between them given by the correlation length of the critical network (see below). Our stable fluctuations pattern reflects a specific distribution of offset charges on the grains, which we call below the background charge distribution. A gate voltage sweep shifts all the offset charges and changes the conductance and/or the position of the dominant single electron transistors.

Annealing the system to room temperature resets the background charge distribution and completely modify the fluctuations pattern as seen in Fig.~\ref{J2CorT}a (a direct evidence of annealing effects in individual grains was seen in Ref.~\cite{KuzminJJAP87}). Our results for annealing at intermediate temperatures (Fig.~\ref{J2CorT}b) may indicate that the background charge distribution is thermally activated. We often observe at 4K a small decrease of the correlation coefficient between V$_g$ scans as time has elapsed, which may correspond to long-term drifts of offset charges at low T.

The absence of periodicity in G(V$_g$) fluctuations could be simply explained by a distribution in C$_g$ values. Significant conductance modulations occur on a scale as small as  $V_g =2$~V, 10 times smaller than the typical scale e/$<$C$_g$$>$. It corresponds to an average change in the charge per grain of only 0.1e. If such a strong sensitivity to V$_g$ changes could also come from a large $C_g$ distribution, we believe that the interactions between electrons on distant grains may play an important role. Indeed, numerical simulations on 1D regular arrays have shown that if the soliton length is large, the threshold voltage is sensitive to gate voltage changes that represent a fraction of electrons per grains \cite{JohanssonPRB00}. Moreover, the modulation pattern of the conductance is strongly dependent on the actual background charge distribution. Unfortunately, measurements of 2D artificial junction arrays made by electronic lithography are not directly comparable to our work \cite{KurdakPRB98,LafargeHab01}. They do not study V$_g$ conductance fluctuations in the linear regime but above the threshold voltage (non ohmic regime) and the interactions are usually short-range ($C/C_g$ below or close to 1).

\subsection{Size and R$_{\Box}$ dependence of the conductance fluctuations}
\label{SizeR}

In order to measure how the fluctuations amplitude depends on the samples size and resistance, we have produced a large set of HM samples. Sizes are distributed between 15~$\mu$m and 1~mm, and R$_{\Box}$ at 4.2~K between 1~M$\Omega$ and 1~G$\Omega$. Unless explicitly mentioned, all the following measurements have been done at 4.2~K. We will show that, in agreement with the previous discussion, most of our results can be interpreted within a general percolation model applied to a system with an exponentially large distribution of microscopic resistances.

\subsubsection{Size dependence of the conductance fluctuations}
\label{SizeDep}

\begin{figure}[h]
\begin{center}\includegraphics[width=8cm]{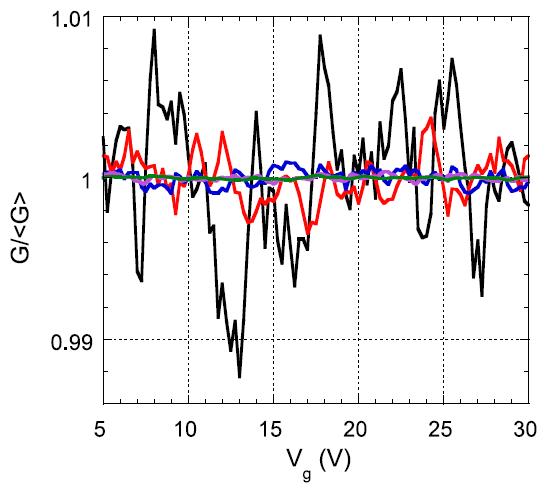}\end{center}
\caption{Normalized conductance for channels of different size (HM, $R_{\Box}=200~M\Omega-1~G\Omega$). Channels with fluctuations amplitudes in decreasing order: $20\mu m \times 15 \mu m$, $50\mu m \times 35 \mu m$, $110\mu m\times 105 \mu m$, $1.1mm\times 50 \mu m$, $1.1mm\times870\mu m$. }
\label{OscilTaille2}
\end{figure}

For samples having similar R$_{\Box}$, the fluctuations amplitude is strongly dependent on the sample size, the larger the channel area, the smaller the amplitude (see Fig.~\ref{OscilTaille2}). Indeed, reproducible fluctuations are still measurable on samples of mm size but the rms relative amplitude falls below 0.01\%.

\begin{figure}[h]
\begin{center}\includegraphics[width=8cm]{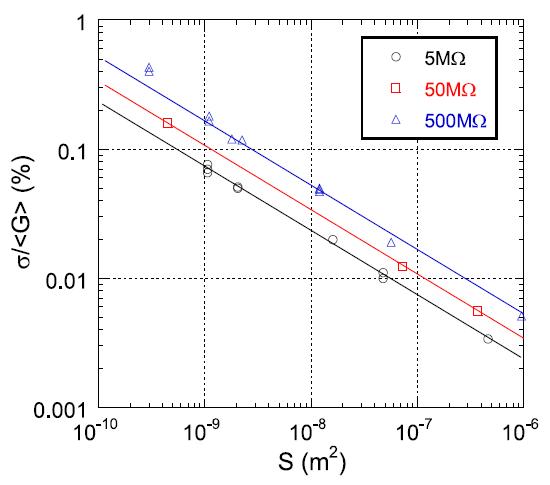}\end{center}
\caption{Relative rms amplitudes of the conductance fluctuations as a function of the channels areas S (HM samples, $T=4.2$~K)). Three different set of samples with similar R$_{\Box}$ are represented: $R_{\Box}\approx 5$~M$\Omega$ (circles), $R_{\Box}\approx 50~$M$\Omega$  (squares) and $R_{\Box}\approx 500~$M$\Omega$ (triangles). The straight lines correspond to the square root dependence of Eq.~\ref{FlucTaille2D} and \ref{FlucTaille3D} (see text for details).}
\label{OscilTaille}
\end{figure}

Relative rms amplitudes are gathered on Fig.~\ref{OscilTaille} for three sets of samples with respectively R$_{\Box}$ about 5~M$\Omega$ , 50~M$\Omega$  and 500~M$\Omega$ . The relative rms amplitude is found to decrease as the invert of the square root of the channel area. This dependence indicates that the observed fluctuations result from the sum of statistically independent microscopic fluctuations. The length scale L$_0$ at which $(G-<G>)/<G>=\delta G /<G> = 1$ is given by

\begin{eqnarray}
  \textrm{if $L_0\ll t$ (2D)}\qquad
  \left(\frac{\delta G}{<G>}\right) &=& \left(\frac{L_0^{\phantom{0}2}}{S}\right)^{1/2}
  \label{FlucTaille2D}\\
  \textrm{if $L_0\gg t$ (3D)} \qquad
  \left(\frac{\delta G}{<G>}\right) &=& \left(\frac{L_0^{\phantom{0}3}}{St}\right)^{1/2}
  \label{FlucTaille3D}
\end{eqnarray}

where t and S are respectively the thickness and the area of the film. Assuming a 2D regime, we get from the data of Fig.~\ref{OscilTaille} $L_0=30$~nm for $R_{\Box}=5$~M$\Omega$  and 60~nm for $R_{\Box}=500$~M$\Omega$. An interpretation of the length L$_0$ comes from the percolation theory applied to strongly inhomogeneous media \cite{ShklovskiiEfros84}. Let's suppose that we have an exponentially wide range of resistances $R_{ij}$ between the sites i and j of an array, i.e. $R_{ij}=\exp(\xi_{ij})$ with an uniform distribution of $\xi_{ij}$ between $0<\xi_{ij}<\Delta\xi$. Then, if $\Delta\xi\gg 1$, the resistance R of the array is that of the critical resistance subnetwork and is given by
\begin{equation}\label{RCritique}
    R=R_0\exp(\xi_C)
\end{equation}
The exponential factor $\xi_C$ is the smallest $\xi_{ij}$ that first gives percolation when only sites such as $\xi_{ij}< \xi_C$ are connected. The critical resistance subnetwork includes all the sites such as  $\xi < \xi_C+1$ (sites with larger $\xi_{ij}$ have a negligible contribution). It is characterized by its correlation radius or homogeneity length L$_0$. In other words, one unavoidable microscopic resistance R$_{ij}$ of the order of the macroscopic one is encountered on a typical length scale L$_0$. L$_0$ is given by
\begin{equation}\label{CorLength}
    L_0\approx l(\Delta \xi)^\nu
\end{equation}  	
$\nu =0.9$ at 3D, $\nu=4/3$ at 2D and $l$ is the microscopic length of the problem (array step size if the hopping is between nearest neighbors or hopping length in the variable range hopping regime). Changing V$_g$ shifts the Fermi level of the system and therefore the values of the microscopic resistances in the system. This gives rise to conductance fluctuations $\delta G/<G>=1$ at the scale L$_0$ and to modifications of the critical resistance network. We will see below whether the condition $\Delta \xi \gg 1$ necessary for the validity of Eq.~\ref{RCritique} and Eq.~\ref{CorLength} are fulfilled in our samples.

The L$_0$  values extracted from data of Fig.~\ref{OscilTaille} are larger than the film thickness (20~nm for HM samples), confirming a 2D-like regime. Our L$_0$ values are significantly smaller than the homogeneity lengths obtained in In$_2$O$_{3-x}$ thin films (L$_0\simeq$ 300~nm) \cite{OrylanchikPRB07}. A study of the fluctuations amplitude dependence on the film thickness (especially around L$_0$) would be very interesting in order to confirm our estimates. The role of the electronic interactions between distant grains on the percolation picture developed above, especially for length scales smaller than the soliton length, should also be clarified.

\subsubsection{R$_{\Box}$ dependence of the conductance fluctuations}
\label{RDep}

\begin{figure}[h]
\begin{center}\includegraphics[width=8cm]{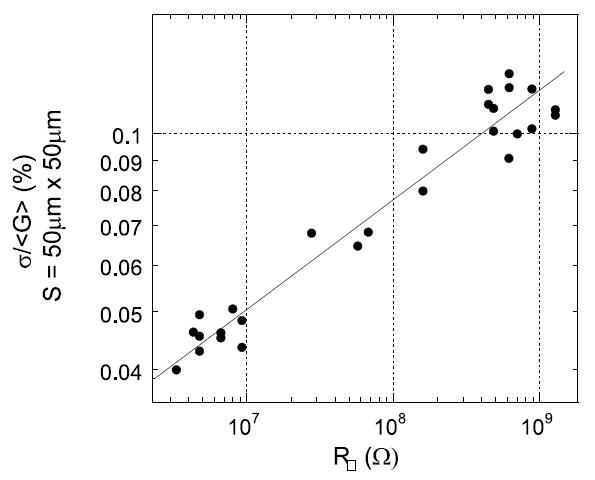}\end{center}
\caption{Relative rms amplitudes of the conductance fluctuations as a function of R$_{\Box}$.The amplitudes correspond to channels area S of $50~\mu m\times50~\mu m$: they have been deduced from data on channels with different sizes and assuming the square root dependence of Eq.~\ref{FlucTaille2D}. The dispersion of the data gives an idea of the rms amplitude uncertainty.}
\label{OscilRsquare}
\end{figure}

 Fig.~\ref{OscilTaille} also emphasizes an increase of the conductance fluctuations amplitude as a function of R$_{\Box}$ for a given channel area. The increase is weak: the relative rms amplitude is multiplied by only 2 when R$_{\Box}$ rises from 5~M$\Omega$  to 500~M$\Omega$  (see also Fig.~\ref{OscilRsquare}).

\begin{figure}[h]
\begin{center}\includegraphics[width=8cm]{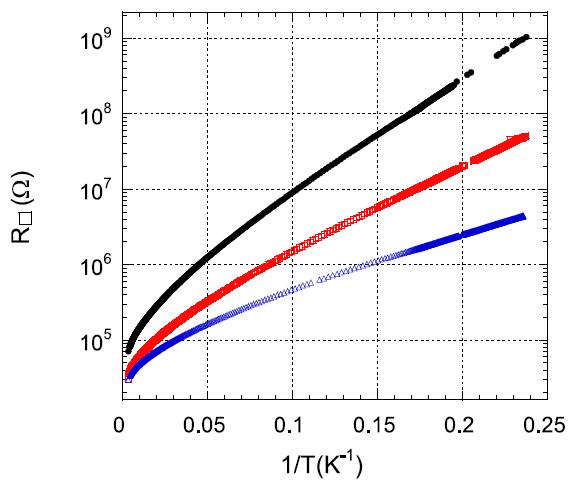}\end{center}
\caption{R$_{\Box}$ as a function of 1/T between 4.2~K and 300~K for three HM samples. The low T part (4.2~K-10~K) could be well described by an activation law. }
\label{RTFigure}
\end{figure}

Below $\simeq10$~K, the resistance temperature dependence is close to an activated law (see Fig.~\ref{RTFigure})
\begin{equation}\label{RTEquation}
    R=R_0\exp(E_0/k_BT)
\end{equation}

\begin{figure}[h]
\begin{center}\includegraphics[width=8cm]{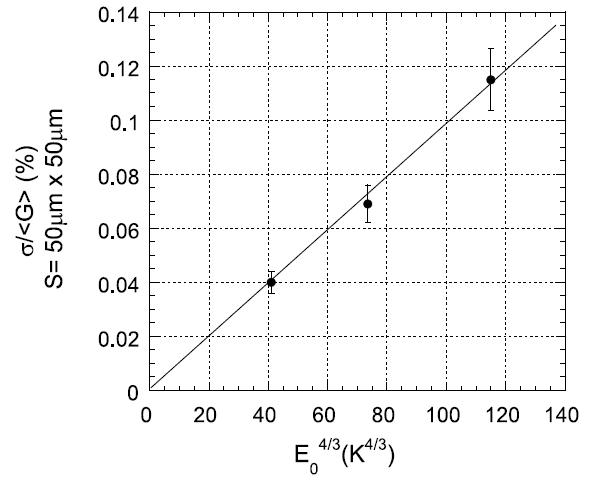}\end{center}
\caption{Relative rms amplitudes of the conductance fluctuations (area S of $50\mu m\times 50\mu m$) as a function of $E_0^{4/3}$. The straight line corresponds to Eq.~\ref{FluctEc}.}
\label{AmpEc}
\end{figure}

According to the percolation model described previously,
\begin{equation}\label{XiEc}
    \xi_C=E_0/k_BT\approx\Delta \xi
\end{equation}
Eq.~\ref{CorLength} and \ref{XiEc} imply that
\begin{equation}\label{L0Ec}
    L_0\approx l(E_0/k_BT)^\nu
\end{equation}
Since for a given channel area the fluctuations amplitude is proportional to L$_0$ (2D-regime) we expect at fixed T:
\begin{equation}\label{FluctEc}
\frac{\delta G}{<G>}\propto E_0^{\phantom 0 \nu}
\end{equation}
In Fig.~\ref{AmpEc}, the rms fluctuations amplitudes are plotted as a function of the activation energies E$_0$ extracted from the fits between 4.2~K and 10~K. The data are in good agreement with Eq.~\ref{FluctEc}, taking $\nu=4/3$. The activation energies range between 16~K and 35~K for HM samples with $R_{\Box}=3$~M$\Omega$ and $1$~G$\Omega$ respectively.
This justify the use of Eq.~\ref{RCritique} and \ref{CorLength} since $\Delta \xi =\xi_C / x_C = E_0/(k_BTx_C)$ is much larger than 1: $\Delta\xi \geq 4$ for $R_{\Box}=3$~M$\Omega$ and $\Delta\xi \geq 8$ for $R_{\Box}=1$~G$\Omega$ ($x_C$, the critical probability, depends on the array type but is always smaller than 1).

%We have also analyzed the data in terms of the Efros and Shklovskii variable range hopping law, a T dependence often found in granular systems
%\begin{equation}\label{ESLaw}
%R=R_{ES}\exp(T_{ES}/T)^{1/2}
%\end{equation}
%In that case, $\Delta\xi \approx (T_{ES}/T)^{1/2}$, $l\approx (T_{ES}/T)^{1/2}$,  and therefore
%\begin{equation}\label{FluctES}
%\frac{\delta G}{<G>}\propto L_0\approx (T_{ES}/T)^{(\nu + 1)/2}
%\end{equation}
%But the low temperature resistance fits and the agreement with Eq.~\ref{FluctES} are less satisfactory than in the case of the activated law analysis.

Up to now, the resistance T dependence of our insulating granular Al thin films is still not fully understood in the whole T range and interactions between electrons may play a role. The activated law of Eq.~\ref{RTEquation} is a reasonable approximation of the low T behavior. Strictly speaking the standard percolation model developed above applies to non-interacting electrons. It is anyway outstanding that the above analysis gives a coherent picture of the data.

\subsection{Temperature dependence of the conductance fluctuations}
\label{TDep}

We have also measured the conductance fluctuations as a function of T above 4K for two samples: one HM sample with $R_{\Box}=1$~G$\Omega$ ($E_0\simeq35$~K) and one EL sample with $R_{\Box}=3$~M$\Omega$ ($E_0\simeq20$~K). A higher T implies a resistance exponentially smaller. In terms of the percolation theory scenario, the correlation length is smaller, the number of dominant microscopic resistances increases in the system which finally reduces the relative fluctuations amplitude. According to Eq.~\ref{L0Ec}, we expect:
\begin{equation}\label{FluctT}
\frac{\delta G}{<G>}\propto (1/T)^\alpha
\end{equation}
With $\alpha=\nu=4/3$ at 2D, $\alpha=(3/2)\nu =1.35$ at 3D.

\begin{figure}[h]
\begin{center}\includegraphics[width=8cm]{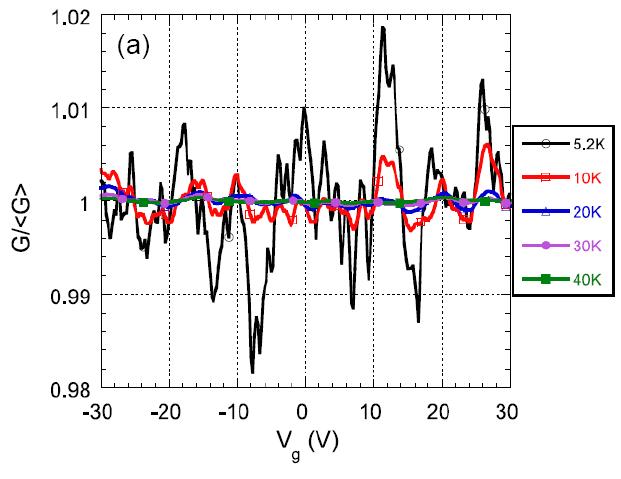}\end{center}
\begin{center}\includegraphics[width=7cm]{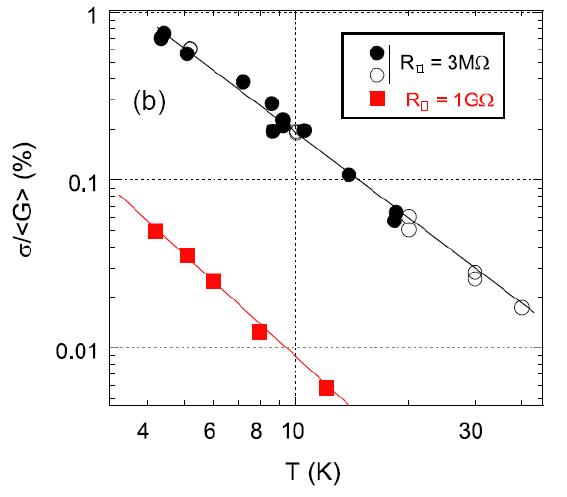}\end{center}
\caption{(a) Relative conductance as a function of V$_g$ for different temperatures. The sample is a channel constriction of $1.8\mu m \times 130nm$ (EL, $R_{\Box}=3$~M$\Omega$). By order of decreasing fluctuations amplitude: $T=5.2$~K, 10~K, 20~K, 30~K and 40~K.
(b) Relative rms fluctuations amplitude as a function of T. Squares: HM sample with channel size $115\mu m \times 105 \mu m$ ($R_{\Box}=1$~G$\Omega$); straight line:  power law Eq.~\ref{FluctT} with $\alpha=2.0$. Circles: EL sample with $R_{\Box}=3$~M$\Omega$  (same channel as in Fig.~\ref{FluctRelT}a); straight line:  power law Eq.~\ref{FluctT} with $\alpha=1.7$. The full and empty circles represent two set of measurements performed on two different cryogenic sticks.}
\label{FluctRelT}
\end{figure}

\begin{figure}[h]
\begin{center}\includegraphics[width=8cm]{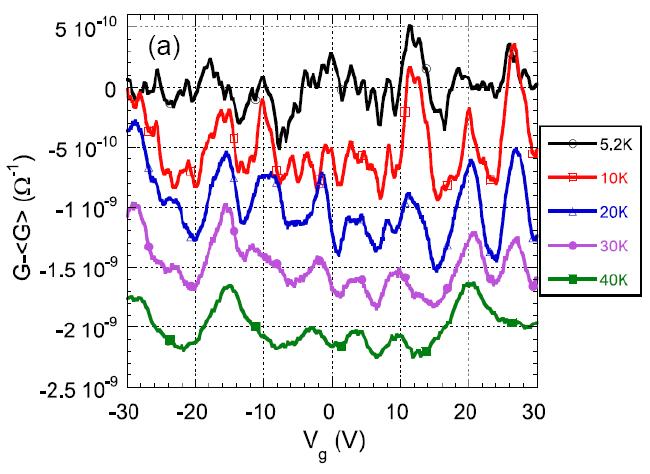}\end{center}
\begin{center}\includegraphics[width=7cm]{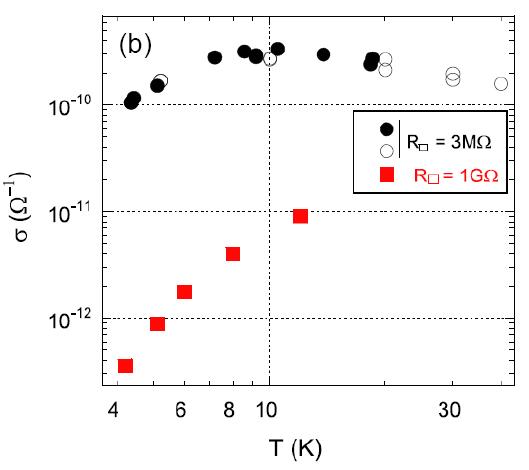}\end{center}
\caption{(a) Absolute conductance variations G$-$$<$G$>$  as a function of V$_g$ for different temperatures. Same sample as in Fig.~\ref{FluctRelT}a. From top to bottom: $T=5.2$~K, 10~K, 20~K, 30~K, 40~K. The lower curves have been shifted for clarity.
(b)~Absolute rms fluctuations amplitude as a function of T. Same samples as in Fig.~\ref{FluctRelT}b.}
\label{FluctAbsT}
\end{figure}

The results for the two samples are presented in Fig.~\ref{FluctRelT} and Fig.~\ref{FluctAbsT}. The agreement with the percolation theory predictions is only qualitative. The relative amplitude decreases as T is increased (see Fig.~\ref{FluctRelT}a). The decrease is well described by the power law Eq.~\ref{FluctT}  but the exponents $\alpha$ are above the expected ones ($\alpha$=2 and 1.7 for respectively the HM and the EL sample, see Fig.~\ref{FluctRelT}b) \cite{TFluctuations}.

As illustrated on Fig.~\ref{FluctAbsT}a and b, the absolute rms fluctuations amplitude first increases and then decreases slightly above $10-20$~K. In Fig.~\ref{FluctAbsT}a, it is seen for the EL sample that the positions of the main structures are unchanged below  20~K (the autocorrelation function width at half value is indeed constant in that T range) and that they are smoothed significantly above. If we still don't know how to explain the observed T dependencies of Fig.~\ref{FluctRelT} and Fig.~\ref{FluctAbsT}, the extension of the measurements down to lower T may bring interesting information on this problem.

\section{The anomalous field effect in presence of the conductance fluctuations}
\label{CondMin}

 One main goal of these experiments was to study the anomalous field effect and its slow relaxations in presence of the conductance fluctuations. Slow conductance relaxation phenomena in our macroscopic (with negligible fluctuations) granular Al films have been described in details in Ref.~\cite{GrenetEPJB03,GrenetEPJB07}.

\subsection{Time evolution of the conductance dip}
\label{CondMinT}

\begin{figure}[h]
\begin{center}\includegraphics[width=8cm]{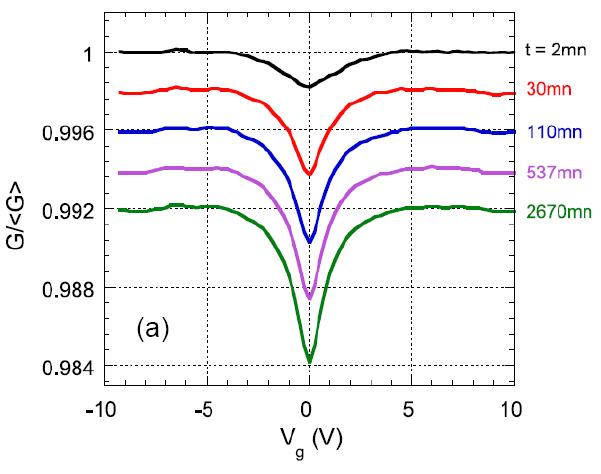}\end{center}
\begin{center}\includegraphics[width=8cm]{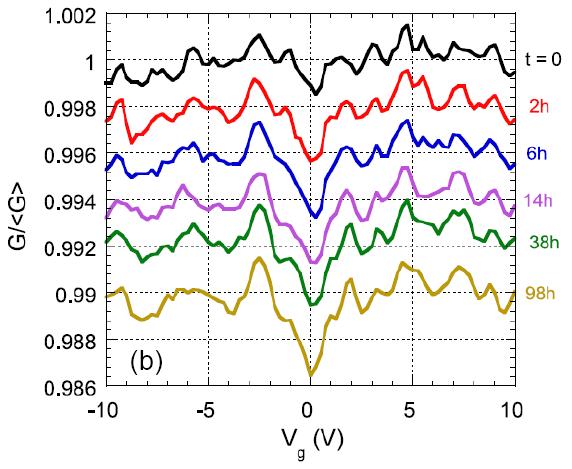}\end{center}
\caption{(a) Normalized conductance for a "macroscopic" sample (HM, $R_{\Box}=30$~M$\Omega$, $100\mu m \times 2 mm$). The sample was quenched from room temperature to 4.2~K at $t=0$ under a fixed gate voltage of $V_{geq}=0$~V. Scans of 240~s long were taken every 30~mn. A symmetric dip centered on V$_{geq}$ increases in amplitude with time. The lower curves have been shifted for clarity.
(b) Normalized conductance for a "microscopic" sample (HM, $R_{\Box}=3.3$~M$\Omega$, channel size $50\mu m\times 40\mu m$). Like in (a), the sample was quenched from room temperature to 4.2~K at $t=0$ under a fixed gate voltage of $V_{geq}=0$~V. Scans of 240~s long were taken every 2 hours. Reproducible conductance fluctuations are superimposed on the conductance dip. The lower curves have been shifted for clarity.}
\label{MacroMicroTrempe}
\end{figure}

In macroscopic samples maintained under a fixed gate voltage V$_{geq}$ at 4.2~K, a fast gate voltage sweep reveals a conductance dip in the G(V$_g$) curves, symmetric and centered on V$_{geq}$. The dip amplitude was found to increase as a logarithm of the time elapsed since $V_g=V_{geq}$. The formation of a conductance dip after a quench at 4.2K under $V_{geq}=0$~V is illustrated in Fig.~\ref{MacroMicroTrempe}a. The same protocol has been followed in Fig.~\ref{MacroMicroTrempe}b but on a smaller sample ($50\mu m\times 40\mu m$ channel). The conductance dip is still present but superimposed on the conductance fluctuations pattern described previously. The two effects are here of equal magnitudes. The striking feature is that the fluctuations pattern is not affected by the digging of the conductance dip, neither for V$_g$ values within the conductance dip range ($\pm 3$~V around V$_{geq}$) nor out of this range. Actually, long-term changes are also visible in the fluctuations of Fig.~\ref{MacroMicroTrempe}b, but they are weak and similar to what is sometimes seen after a quench in samples where the conductance dip is not visible. Since the fluctuations and the noise amplitude increases when the sample area is reduced (which is not the case for the dip amplitude, see Sect.~\ref{CondMinPara}), the conductance dip is not visible in too small samples (channels size below $\simeq 10\mu m)$. This was for example the case for the EL sample pattern of Fig.~\ref{J2ChgtVg}.

\begin{figure}[h]
\begin{center}\includegraphics[width=8cm]{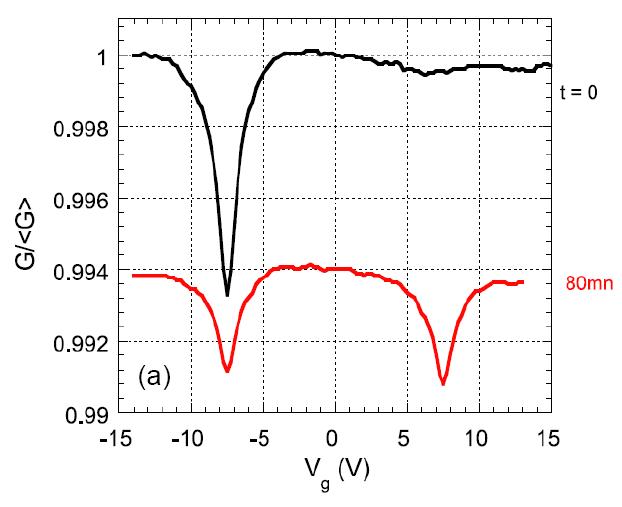}\end{center}
\begin{center}\includegraphics[width=8cm]{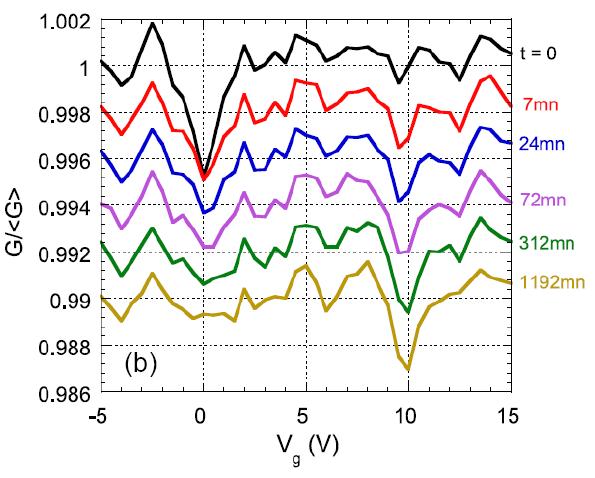}\end{center}
\caption{(a) Normalized conductance for a "macroscopic" sample (same sample as in Fig.~\ref{MacroMicroTrempe}a). The sample was maintained 24h under $V_g=-7.5$~V and the gate voltage was changed to +7.5~V after the scan labeled $t=0$~s. 80~mn later, a new dip is present at 7.5~V while the amplitude of the dip at -7.5~V has been reduced. The lower curve has been shifted for clarity.
(b) Normalized conductance for a "microscopic" sample (same sample as in Fig.~\ref{MacroMicroTrempe}b). The sample was maintained 4 days under $V_{geq1}=0$~V and the gate voltage was changed to $V_{geq2}=+10$~V after the scan labeled $t=0$. A new dip which amplitude increases with time is formed at +10~V while the "old" dip at 0~V is erased with time. Scans are 80~s long, with 400~s between two scans. The lower curves have been shifted for clarity.}
\label{MacroMicroChgtVg}
\end{figure}

We have also tested a second procedure. Once the sample has been kept for a long time (days) under V$_{geq1}$, the gate voltage is changed to a different value V$_{geq2}$. The results are illustrated for a macroscopic sample in Fig.~\ref{MacroMicroChgtVg}a: a new dip is formed at V$_{geq2}$ while the old dip at V$_{geq1}$ is slowly erased. In microscopic samples (see Fig.~\ref{MacroMicroChgtVg}b), the conductance dip behaves similarly and the fluctuations pattern is essentially unchanged under the displacement of the conductance dip.

\begin{figure}[h]
\begin{center}\includegraphics[width=8cm]{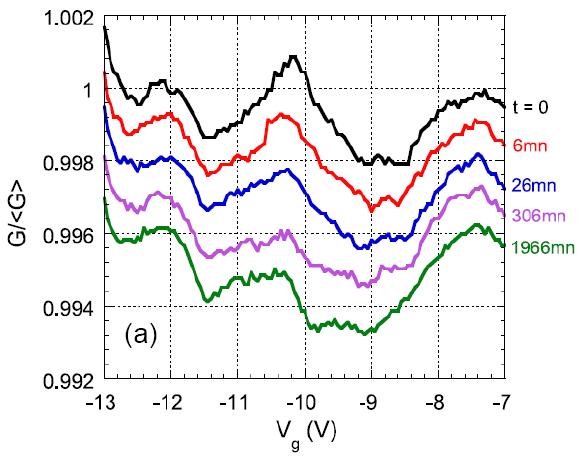}\end{center}
\begin{center}\includegraphics[width=8cm]{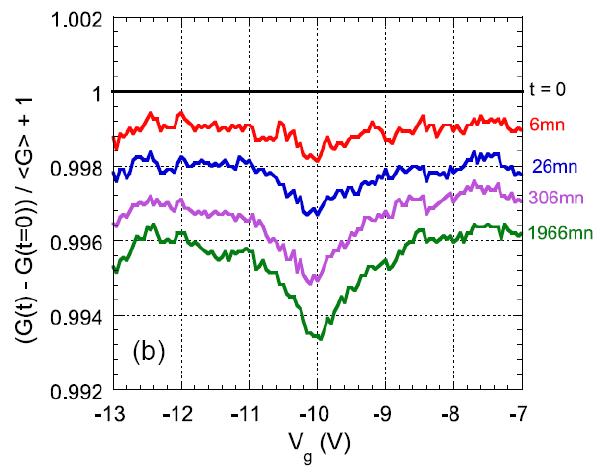}\end{center}
\caption{(a) Normalized conductance for a "microscopic" sample (same sample as in Fig.~\ref{MacroMicroChgtVg}b). The sample was maintained three days under $V_{geq1}=0$~V before the gate voltage was changed to $V_{geq2}=10$~V after the scan labeled $t=0$. The V$_{g}$ values of the sweeps focus on the dip range.
(b) Difference between the $G(V_g)$ curves at time $t>0$ (measured  under V$_{geq2}$) and at $t=0$ (measured under V$_{geq1}$).}
\label{ZoomMicroChgtVg}
\end{figure}

This is even clearer in  Fig.~\ref{ZoomMicroChgtVg}a and b, where similar data are presented for a microscopic sample but on a reduced V$_g$ range around V$_{geq2}$. By subtracting the curve at $t=0$ (measured under V$_{geq1}$) from the other curves (measured under V$_{geq2}$), the stable fluctuations pattern is eliminated and we are let with a smooth and symmetrical conductance dip (see Fig.~\ref{ZoomMicroChgtVg}b).  Thus, like for the previous procedure, the fluctuations pattern is not significantly modified by the formation or the erasing of a conductance dip, for gate voltages inside or outside the conductance dip range. We have carefully checked this point in many samples having different R$_{\Box}$ values and down to an area of $20\mu m \times 15\mu m$.

\subsection{Parameters of the conductance dip: width and amplitude}
\label{CondMinPara}

\begin{figure}[h]
\begin{center}\includegraphics[width=8cm]{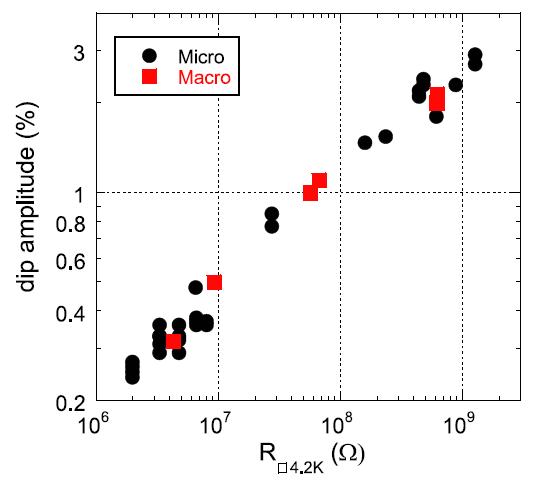}\end{center}
\caption{Relative amplitude of the conductance dip for different HM samples. The amplitudes correspond to that of dips dug for one hour at 4.2~K (see text for details). The microscopic (circles) and macroscopic (squares) samples have respectively areas in the range $10^{-10}-10^{-8}$~m$^2$ and $10^{-8}-10^{-6}$~m$^2$.}
\label{Amp1hR}
\end{figure}

We have compiled conductance dip amplitude data measured on samples of different sizes (lowest area $20\mu m\times 15\mu m$) and different R$_{\Box}$ values. Since the dip is an off-equilibrium property, it is crucial to use exactly the same procedure for all the measurements. In Fig.~\ref{Amp1hR}, the parameters are the following: the samples are first maintained under a fixed V$_{geq1}$ for at least 24~h. The gate voltage is then changed to V$_{geq2}$ for 1 hour, creating a new dip, and a gate voltage scan 20~V wide starting from V$_{geq2}$ is recorded (81 points, 4~s/point). The difference between the conductance out of the dip range and the conductance at V$_{geq2}$ gives the new dip amplitude. For microscopic samples (where significant fluctuations are present), a reference scan taken before the change to V$_{geq2}$ is subtracted.  The results are independent of the V$_{geq1}$ and V$_{geq2}$ values. Fig.~\ref{Amp1hR} clearly demonstrates that, in spite of a larger data dispersion in microscopic samples (due essentially to a larger noise level), the conductance dip amplitude depends on R$_{\Box}$ and not on the sample area in contrary to the fluctuations amplitude. Moreover, the R$_{\Box}$ dependence of the dip amplitude is faster than that of the conductance fluctuations. Indeed, its relative value is multiplied by 10 between 2M$\Omega$  and 1G$\Omega$ , instead of 3 for the fluctuations amplitude (see Fig.~\ref{OscilRsquare}).

\begin{figure}[h]
\begin{center}\includegraphics[width=8cm]{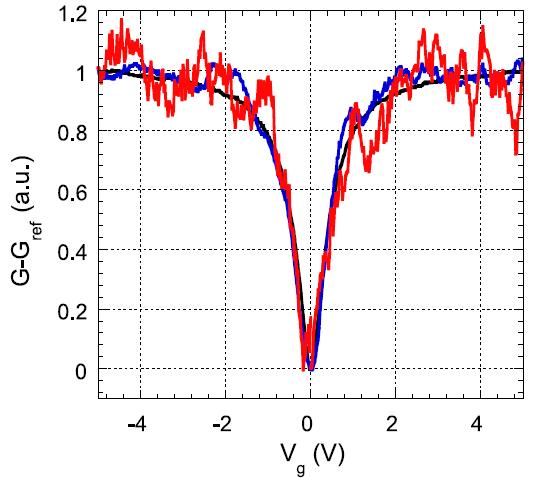}\end{center}
\caption{Shapes of the conductance dip for HM samples of different sizes. The stable fluctuations have been removed by subtracting a reference $G_{ref}(V_g)$ scan taken before the formation of the dip. The fluctuations that remain (larger in smaller samples) are due to the short time noise level of the scans. Channels sizes are: $1.1mm\times50\mu m$ ($R=60$~M$\Omega$ ), $50\mu m\times 40\mu m$ ($R=6.6$~M$\Omega$ ), and $20m\mu\times15\mu m$ ($R=240$~M$\Omega$ ). The conductance was arbitrarily set to 0 in its minimum and to 1 out of the dip range.}
\label{LargeurTaille}
\end{figure}

The conductance dip width is also the same in the different samples we have measured, independently of the conductance fluctuations amplitudes and the R$_{\Box}$ values (see Fig.~\ref{LargeurTaille}). As it was already mentioned in Ref.~\cite{GrenetEPJB07}, the width of the dip in our granular Al films is fixed by the temperature of the measurement and is independent of R$_{\Box}$ values and V$_g$ sweep parameters.

\subsection{Time relaxation of the conductance}
\label{TimeRelax}

\begin{figure}[h!]
\begin{center}\includegraphics[width=8cm]{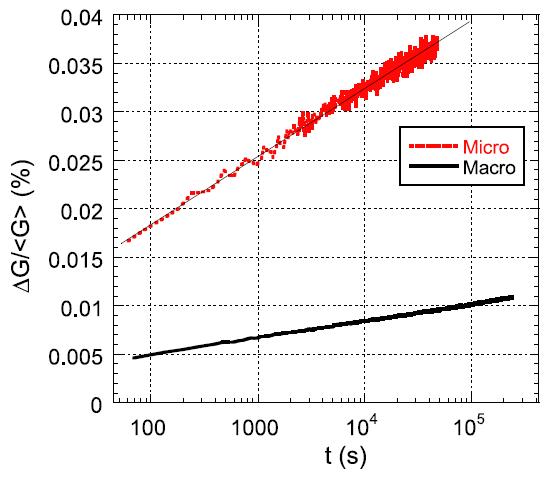}\end{center}
\caption{Relative amplitude of a new conductance dip a time t after the gate voltage is changed from V$_{geq1}$ to V$_{geq2}$ [$\Delta G =<G>-G(V_{geq2}$)] for a macroscopic (lower curve) and a microscopic (upper curve) samples. Macroscopic sample: HM, $R_{\Box}=30M\Omega$ , channel size $100\mu m\times 2mm$. Microscopic sample: HM, $R_{\Box}=450M\Omega$ , channel size $45\mu m\times 50\mu m$. The V$_{geq2}$ dip amplitudes increase as a logarithm of time (straight line). The values are different for the two samples according to their different $R_{\Box}$ values.}
\label{NewDip}
\end{figure}

Finally, we have tested if the conductance relaxation laws were modified by the presence of conductance fluctuations. One sample is maintained under a fixed gate voltage during a few days and we follow the conductance response to a sudden change of V$_g$. In macroscopic samples \cite{GrenetEPJB07}, after a fast increase, the conductance decreases as a logarithm of the time elapsed since V$_g$ was changed. In presence of conductance fluctuations, the decrease is noisier but is still described by a logarithm of time (see Fig.~\ref{NewDip}).

\begin{figure}[h!]
\begin{center}\includegraphics[width=8cm]{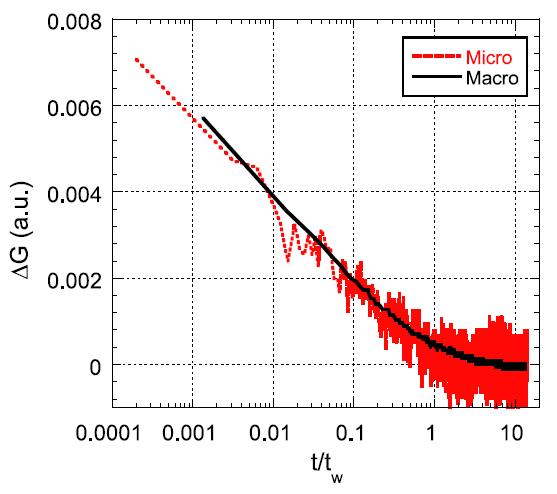}\end{center}
\caption{Two dips experiment measurement in a macroscopic and a microscopic samples. We plot how the $V_{geq2}$ dip is erased with time once $V_g$ is switched back to $V_{geq1}$ [$\Delta G = <G>-G(V_{geq2})$, see text for details]. Macroscopic sample: HM, $R_{\Box}=30$~M$\Omega$ , channel size $100\mu m\times 2mm$, $t_w=3600$~s. Microscopic sample: HM, $R_{\Box}=3.3$~M$\Omega$, channel size $50\mu m\times 40\mu m$, $t_w=10000$~s. The microscopic relaxation curve has been shifted and normalized on the macroscopic one for comparison.}
\label{2Dips}
\end{figure}

We have also tested the so called two dips experiment \cite{VakninPRB02}. In this protocol, the sample is maintained for a long time under V$_{geq1}$, a new dip is formed at V$_{geq2}$ during a time t$_w$ and the evolution of $G(V_g)$ curves is followed when the gate voltage is switched back to V$_{geq1}$. At least when $\mid V_{geq1}-V_{geq2}\mid$ is not too large and in macroscopic samples, the erasing of the V$_{geq2}$ dip amplitude was found to be a universal function of $t/t_w$ (simple aging) \cite{VakninPRB02,GrenetEPJB07}. Once again, we have observed the same relaxation curves in macroscopic and microscopic samples (see Fig.~\ref{2Dips}), whatever the conductance fluctuations amplitude.

\subsection{Discussion}
\label{Discussion}

Our measurements demonstrate the following important features:
\begin{itemize}
  \item[-] the glassy conductance dip is not significantly altered by the samples size down to 20~$\mu$m. This suggests that the conductance dip doesn't result from an incomplete self-averaging of random microscopic conductance modulations. Its shape and amplitude are indeed independent of the sample size even when the fluctuations amplitude is of the same order of magnitude. The uncertainty in the conductance dip parameters increases when the sample size is reduced, but this increase comes mainly from a higher noise level
 \item[-] the conductance fluctuations and the dip seem to be independent from each other. Indeed, the fluctuations pattern is not affected by the formation of the conductance dip, and conversely, the conductance dip is formed whatever the conductance fluctuations amplitude. If we assume that the fluctuations pattern reflects the specific potential landscape of the grains, then we conclude that the formation of the conductance dip does not induce large changes in this potential landscape
\item [-]the conductance fluctuations and the dip have different dynamics: the former are essentially static once the sample has been quenched at 4~K, while the dip slowly grows. As we have seen, a dynamics of the fluctuations is induced by heating the samples, while no signature of thermal activation can be observed in the glassy dynamics of the dip \cite{GrenetEPJB07}.
\end{itemize}
The independence of the conductance dip and the conductance fluctuations was also demonstrated recently on In$_2$O$_{3-x}$ thin films down to channel sizes of 2~$\mu m$ and with a maximum relative amplitude of the minimum about 10\% \cite{OrylanchikPRB07}. Our results underline once again the strong similarities between granular Al and indium-oxide.

What conclusions may be drawn from these experimental observations about the origin of the glassy dip? In a simplified intrinsic scenario, the stable fluctuations pattern reflects the static distribution of charges in the dielectric material around the grains whereas the slow conductance decrease (dip formation at constant V$_g$) comes from slow correlated electron hops. This scenario would naturally explain why the two phenomena are independent and have different dynamics, so that our observations are in qualitative agreement with an electron glass interpretation of the conductance dip. But this simplified picture is certainly not exact in our system. As already discussed in Sect.~\ref{Interpretation} for a regular array of tunnel junctions, trapped charges in the dielectric materials around the grains result in a random distribution of offset charges between $-e/2$ and $e/2$ only when the grains are independent ($C\ll C_g$). However, this distribution does not minimize the global electrostatic energy when $C>C_g$ due to interactions between the electrons on distant grains. Tunneling of single electrons between the grains permits a large and fast diminution of the electrostatic energy of the system \cite{KaplanPRB03}. The random potential is strongly modified by these fast relaxation processes: the potential landscape is smoothed \cite{JohanssonPRB00}, strengthening the Coulomb blockade and opening a Coulomb gap in the density of states of single electron addition energies on the grains \cite{KaplanPRB03}. Thus the random potential (and the conductance fluctuations) are partly of electronic origin and cannot be considered decoupled from the electronic degrees of freedom. The conductance dip formation is attributed to the slower part of the electronic degrees of freedom, consisting of correlated electrons hops \cite{LebanonPRB05}. Correlated electron hops are still decreasing the global electrostatic energy but they are not able to change significantly the local electrostatic order. As long as the conductance dip relative amplitude is small, it seems reasonable to assume that the slow correlated electron hops only induce small shifts in the grains potential (or possibly rare large shifts). For a given V$_g$, the critical resistance subnetwork is therefore essentially fixed by the trapped charges in the dielectric material around the grains and by the fast single electrons hops between the metallic grains. In other words, we expect that the conductance dip formation affects only slightly the conductance fluctuations pattern. The apparent observed independence of the two phenomena may thus simply come from the fact that the conductance dip amplitude is still small. To this respect, similar measurements on samples having larger conductance dip amplitudes would be of utmost interest.

According to the extrinsic scenario, the conductance dip formation is the result of a slow response of the background charges around the grains. In a continuous model, this is described as a slow linear polarization of the dielectric surrounding the charged grains. It could be shown by a subtle argument that such a polarization shifts the grains potentials in a way that decreases the electron mobility \cite{CavicchiPRB88}. The model was initially developed for an electron transport perpendicular to a 2D-array of independent grains: starting from an uniform distribution of offset charges between $-e/2$ and $+e/2$, the polarization completely redistributes the offset charges values, the final effect being to reduce the proportion of grains with offset charges close to $+e/2$ or $-e/2$ (the "unblocked" grains). Even if a detailed theory for the transport along a 2D-array of interacting grains is still missing, we think that the polarization could similarly reduce the proportion of low resistance hops in our granular films \cite{GrenetEPJB07}. Since the dip amplitude is small and since it doesn't result from the incomplete self-averaging of a random and strongly fluctuating microscopic quantity, it is reasonable to assume that the part of the polarization responsible for the conductance dip induces either small shifts in the grains potentials or larger shifts but for a small number of grains. If such shifts are triggered off by changes in the charge configuration of two level systems nearby, we believe that the second hypothesis is more realistic. According to Sect.~\ref{Fluctuations}, the conductance fluctuations are instead associated with large changes in the background charge distribution and consequently in the critical resistance subnetwork. We could then naturally understand why the rare events associated with the formation of a small conductance dip could not influence significantly the fluctuations pattern. However it is more difficult to explain the difference in the conductance dip and fluctuations dynamics. A more quantitative approach of the extrinsic scenario (e.g. via numerical simulations) would be very helpful to see whether it can indeed reproduce the coexistence of the dip and the fluctuations.

\section{Conclusion}
\label{Conclusion}

In conclusion, we have observed at 4~K reproducible conductance fluctuations as a function of gate voltage in insulating granular Al thin films of micrometer size. Such fluctuations are stable over days of measurement. Their amplitude depends on the size and the resistance per square of the films in agreement with a percolation model applied to an exponentially wide distribution of microscopic resistances. Such wide distribution comes from the combined effects of a charging energy larger than the thermal energy and an important potential disorder of the grains. To the accuracy of our measurements, the conductance fluctuations pattern is not affected by the slow formation of the conductance dip. Conversely and down to sizes of $\simeq20$~$\mu m$, the conductance dip parameters are unchanged even when the two phenomena have the same amplitudes. This apparent independence may result mainly from the fact that the conductance dip amplitude corresponds to small changes of the macroscopic conductance (no more than few percent in our case). Our results may be more readily interpreted in the intrinsic context (electron glass origin of the conductance dip), but further theoretical analysis would be needed to definitely rule out the extrinsic scenario.

\section{Acknowledgements}

We acknowledge M. M\"uller for fruitful discussions. We also thank Sylvain Dumond and the personnel of NANOFAB services for technical support. This research has been partly supported by the French National Research Agency ANR (contract N$^{\circ}$ANR-05-JC05-44044)

% BibTeX users please use
% \bibliographystyle{}
% \bibliography{}

\begin{thebibliography}{}
\bibitem{CorCoeff}
The linear correlation coefficient $C_{12}$ between two conductance fluctuations patterns $G_1(V_{gi})=G_{1i}$ and $G_2(V_{gi})=G_{2i}$ with i=1,...N was calculated according to the usual formula: $C_{12}=\frac{N\sum G_{1i} G_{2i}-\sum G_{1i}\sum G_{2i}}{\sqrt{N\sum \left(G_{1i}\right)^2-\left(\sum G_{1i}\right)^2}\sqrt{N\sum \left(G_{2i}\right)^2-\left(\sum G_{2i}\right)^2}}$
\bibitem{note2}
At room temperature, the Coulomb blockade is inefficient and the sample resistance $R_{\Box}$ is fixed by the distribution of the microscopic tunnel resistances {$R_{ti}$} alone. If this distribution width is small ($R_{ti}\simeq R_t$) and asuming that we have 4 grains in the thickness of the film, $R_t\simeq 4R_{\Box}$. For a larger distribution, some tunnel resistances would not contribute to the $R_{\Box}$ value (too large or too small $R_{ti}$) and dominant $R_{ti}$ values stand between $ R_{\Box}$ (2D case, very large distribution width) and $ 4R_{\Box}$.
\bibitem{TFluctuations}
There are many problems with the use of the percolation theory and Eq.~\ref{FluctT} in our samples. First, the T dependence of the resistance is not really activated above 10K (see Fig.~\ref{RTFigure}). Second, the percolation theory results are only valid if $\Delta\xi$ is much larger than 1 \cite{ShklovskiiEfros84}, which may be the case in both samples at 4K ($E_0/k_BT>5$) but certainly not at 30~K for the EL sample ($E_0/k_BT \simeq 0.7$). Finally, estimates for L$_0$ at 10~K in the HM sample gives about 10~nm (with the use of the 3D Eq.~\ref{FlucTaille3D}), which is quite close to the grains diameter.
\end{thebibliography}

\end{document}